\documentclass[a4paper,11pt]{article}
\pdfoutput=1 
\usepackage[utf8]{inputenc}
\usepackage{jheppub} 
\usepackage[T1]{fontenc} 
\graphicspath{{img/}} 

\title{\boldmath Charge imbalance resolved R\'enyi negativity for free compact boson: Two disjoint interval case}

 \author{Himanshu Gaur}
 \author{and Urjit A. Yajnik,}
 \affiliation{Department of Physics, Indian Institute of Technology Bombay, Powai, Mumbai, Maharashtra 400076 India}

\emailAdd{194123018@iitb.ac.in}
\emailAdd{yajnik@iitb.ac.in}

\abstract{In this paper, we study the symmetry decomposition of R\'enyi negativity into charge imbalance sectors for the 1+1 dimensional free compact boson field with a global U(1) symmetry in the ground state for the case of two disjoint intervals. We obtain multi-charged and charged R\'enyi negativity moments by computing the four-point correlator of flux-generating vertex operators on the Riemann surface. We then obtain charge imbalance resolved R\'enyi negativity by taking the Fourier transform of the charged moments. Finally, we match our results against the tight-binding model as a numerical check.}

\begin{document} 
\maketitle
\flushbottom

\section{Introduction} \label{introduction}
The utility of entanglement in quantum theory cannot be overemphasised, it has proven to be an essential tool in the study of black hole entropy \cite{a1}, gauge/gravity duality \cite{a2,a3}, quantum computation \cite{a4}, and criticality in quantum many-body systems \cite{a5}. In quantum many-body systems, entanglement shows scaling behaviour near the quantum critical points \cite{a5,a6,a7}. Among the variety of entanglement measures, entanglement entropy and R\'enyi entropy are the most prominent measures for the pure states. To study entanglement, we take the system to be in a pure state $|{\Psi}\rangle$ and partition it into two subsystems $A$ and its complement $B$, such that the Hilbert space is $\mathcal{H}=\mathcal{H}_A\otimes\mathcal{H}_B$. The reduced density matrix $\rho_A$ for the subsystem $A$ is given by $\mathrm{Tr}_B|{\Psi}\rangle\langle{\Psi}|$. Entanglement entropy $S^A_1$ and R\'enyi entropy $S_n^A$ are given by
\begin{align}
S^A_1&=-\mathrm{\rho_A\ln \rho_A} \label{eq:1.1},\\
S_n^A&=\frac{1}{1-n}\ln\mathrm{Tr}\rho_A^n \label{eq:1.2}.
\end{align} 
Entanglement entropy for a single spatial interval in the ground state of a $1d$ critical theory is proportional to the central charge $c$ of the theory and scales as the logarithm of the subsystem length \cite{a8,a9,a10}. In the present study, we further partition $A$ into $A_1$ and $A_2$ (such that $\mathcal{H}_A=\mathcal{H}_{A_1}\otimes\mathcal{H}_{A_2}$) with the aim to study entanglement between these two subsystems. In this case, both these entanglement measures fail since we have a mixed density matrix for the subsystem $A_1\cup A_2$. To characterise entanglement for a bipartite in the case of mixed states, negativity measures are particularly useful. Two effective measures for the subsystem $A_1\cup A_2$ are log negativity $\mathcal{E}$ and negativity $\mathcal{N}$ \cite{a11,a12,a5}
\begin{align} \label{eq:1.3}
\mathcal{E}&=\mathrm{Tr}\left(\ln\left|\rho_A^{T_2}\right|\right),\\
\mathcal{N}&=\frac{1}{2}\left(\mathrm{Tr}|\rho_A^{T_2}|-1\right),
\end{align}
where $\rho_A^{T_2}$ is obtained by taking the partial trace of $\rho_A$ over $A_2$ degrees of freedom. However, for large systems $\ln\left|\rho_A^{T_2}\right|$ is not easily computable. To circumvent this problem, the replica trick is adapted for negativity measures. We introduce R\'enyi negativity
\begin{equation}
R_n\equiv\mathrm{Tr}\left(\rho_A^{T_2}\right)^{n}.
\end{equation} 
Log negativity and negativity may then be obtained from $R_n$ via
\begin{align} \label{eq:1.4}
\mathcal{E}&=\lim_{n_e\to 1}\ln R_{n_e}, \\
\mathcal{N}&=\lim_{n_e\to 1}\frac{1}{2}(R_{n_e}-1),
\end{align}
where $n_e$ implies that we analytically continue the replicas obtained for even integer values of $n$ to non-integer values. The knowledge of $R_n$ can be exploited to determine the negativity spectrum \cite{a14}. $R_n$ also naturally encapsulate many properties of log negativity and negativity. Negativity measures have been substantially investigated in critical theories \cite{a14,a15,a16,a17,a18,a19,a19i,a52,a20}.

When a system possesses a global internal symmetry that is additive in the subsystems, the entanglement and the negativity measures discussed above decompose into the local charge sectors for the states with fixed global charge corresponding to the global internal symmetry. The study of entanglement in these charge sectors has been termed symmetry resolution of entanglement \cite{a21,a22}. Recently there has been substantial research in symmetry resolved entanglement \cite{a23,a24,a25,a26,a27,a28,a29,a29i,a30,a30i,a30ii,a30iii}, and symmetry resolved negativity \cite{a31,a31i,a32,a33,a34} for quantum many-body systems with a $U(1)$ symmetry. For higher symmetry groups, the Wess-Zumino-Witten models have been studied in a similar context \cite{a35}. Symmetry resolved entanglement in the context of AdS/CFT correspondence has been studied in ref. \cite{a36,a37,a37i,a38}. Experimental protocols for the detection of symmetry resolved entanglement in quantum many-body systems have been proposed for pure and mixed states in ref. \cite{a22,a38i}.

\begin{figure}
\centering 
\includegraphics[width=0.8\textwidth]{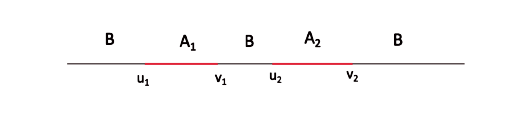}
\caption{\label{fig:i} Two disjoint intervals, intervals $A_1$ and $A_2$ are $(u_1,v_1)$ and $(u_2,v_2)$ respectively, here $\ell_1=|v_1-u_1|$, $\ell_2=|v_2-u_2|$ and $d=|u_2-v_1|$}
\end{figure}

In this work, we study the symmetry resolution of the R\'enyi negativity for the 1+1 dimensional free compact boson for the case where $A_1$ and $A_2$ are two disjoint intervals. The free compact boson is a conformally invariant field with a $U(1)$ symmetry and describes the Luttinger liquids. The set-up for our study is shown in figure \ref{fig:i}, it consists of two intervals $A_1$, and $A_2$ of length $\ell_1$, and $\ell_2$ respectively. The distance between the two intervals is denoted by $d$.

The organisation of this paper is as follows. In section \ref{sre}, we briefly discuss the symmetry decomposition of entanglement in the case of $U(1)$ symmetry. In section \ref{sec3}, we discuss the replica trick in conformal field theory to find the charged moments. In section \ref{section4}, we briefly introduce the theory of free compact boson and find the charged moments for the R\'enyi negativity. We also numerically verify our results against the tight-binding model. In section \ref{section5}, we find the charged imbalance resolved R\'enyi negativity and numerically verify our results here as well. In section \ref{section6}, we discuss and conclude our work. In appendices \ref{A1}, \ref{A2}, and \ref{A4}, some necessary computations are discussed. Finally, in appendix \ref{A5}, we revisit the charged moments for the complex bosons.
\section{Symmetry decomposition} \label{sre}
In this section, we review the symmetry decomposition of the R\'enyi entropy \cite{a21} and the R\'enyi negativity \cite{a22} in a theory with a global $U(1)$ symmetry.

We consider a bipartite, subsystem $A$ and its complement $B$. We assume that the charge $\hat{Q}$ associated with the global $U(1)$ symmetry has a local decomposition into $A$ and $B$ i.e. $\hat{Q}=\hat{Q}_A+\hat{Q}_B$. For the system in a state with a fixed global charge, we have $[\rho,\hat{Q}]=0$. Taking the trace over the degrees of freedom in $B$, we obtain $[\rho_A,\hat{Q}_A]=0$. This implies that $\rho_A$ is block diagonal in the local charge sectors. We may study the R\'enyi entropy in the local charge sector characterised by eigenvalue $q$ of the local charge $\hat{Q}_A$ by taking the trace of the $n^{th}$ power of the block $\rho_{A,q}$ corresponding to $q$. However, block decomposition does not always readily manifest. To circumvent this problem we introduce the charged moments 
\begin{equation} \label{eq:2.1}
Z_n\left(\alpha\right)=\mathrm{Tr}\left(\rho_A e^{i\alpha\hat{Q}_A}\right).
\end{equation}
The symmetry resolved R\'enyi entropy $S_{n,q}$ may be computed after evaluating the Fourier transform of the charged moments
\begin{align} \label{eq:2.2}
\mathcal{Z}_{n}(q)&=\int_{-\pi}^\pi\mathrm{d}\alpha e^{-i\alpha q}Z_n\left(\alpha\right),\\
S_{n,q}&=\frac{1}{1-n}\ln\left( \frac{\mathcal{Z}_{n}(q)}{\mathcal{Z}_{1}(q)^n}\right).
\end{align}
In the case where $A=A_1\cup A_2$, we further assume that the charge $\hat{Q}_A$ has a local decomposition in $A_1$ and $A_2$ i.e. $\hat{Q}_A=\hat{Q}_{A_1}+\hat{Q}_{A_2}$. In this case, we may generalise the charged moments to
\begin{equation} \label{eq:2.3}
Z_n\left(\alpha,\beta\right)=\mathrm{Tr}\left(\rho e^{i\alpha\hat{Q}_{A_1}+i\beta\hat{Q}_{A_2}}\right).
\end{equation}
$Z_n\left(\alpha,\beta\right)$ has been termed multi-charged moments and was first introduced in ref. \cite{a29}. After taking the Fourier transform, we obtain the quantities $\mathcal{Z}_n^{A_1:A_2}(q_1,q_2)$, where $q_1$ and $q_2$ are eigenvalues of the charge operators $Q_{A_1}$ and $Q_{A_2}$ respectively
\begin{equation} \label{eq:2.4}
\mathcal{Z}_n^{A_1:A_2}(q_1,q_2)=\frac{1}{(2\pi)^2}\int_{-\pi}^\pi\mathrm{d}\alpha\int_{-\pi}^\pi\mathrm{d}\beta\, e^{-i\alpha q_1-i\beta q_2}Z_n\left(\alpha,\beta\right).
\end{equation} 
The quantity $\mathcal{Z}_1^{A_1:A_2}(q_1,q_2)$ is interpreted as the joint probabilities of getting $q_1$ and $q_2$ from measurement of $\hat{Q}_{A_1}$ and $\hat{Q}_{A_2}$ respectively.

Negativity measures involve taking the partial transpose of the subsystem $A_2$. In this case, we have the relation
\begin{equation} \label{eq:2.5}
[\rho^{T_2},\mathcal{\hat{Q}}]=0,\hspace{0.2in} \text{where}\hspace{0.1in} \mathcal{\hat{Q}}=\hat{Q}_{A_1}-\hat{Q}_{A_2}^{T}.
\end{equation}
$\mathcal{\hat{Q}}$ are known as the charge imbalance operator. In order to resolve the R\'enyi negativity into the charge sectors of $\mathcal{\hat{Q}}$, called charge imbalance resolved R\'enyi negativity, we introduce the charged R\'eyni negativity moments
\begin{equation} \label{eq:2.6}
R_n(\alpha)=\mathrm{Tr}\left[\left(\rho_A^{T_2}\right)^n e^{i\mathcal{\hat{Q}}\alpha}\right].
\end{equation}
The R\'enyi negativity $\mathcal{R}_n(q)$ in the charge sector $q$, where $q$ are the eigenvalues of $\mathcal{\hat{Q}}$, is obtained by taking the Fourier transform of the charged moments
\begin{equation} \label{eq:2.7}
\mathcal{R}_{n}(q)=\int_{-\pi}^\pi\mathrm{d}\alpha\,e^{-i\left(q-\langle\mathcal{\hat{Q}}\rangle\right)\alpha}R_n(\alpha).
\end{equation}
We may also introduce multi-charged moments for the R\'enyi negativity
\begin{equation} \label{eq:2.8}
R_n\left(\alpha,\beta\right)=\mathrm{Tr}\left(\rho_A e^{i\alpha\hat{Q}_{A_1}-i\beta\hat{Q}_{A_2}^{T}}\right),
\end{equation}
where $q_1$ and $q_2$ are eigenvalues for $\hat{Q}_{A_1}$ and $-\hat{Q}_{A_2}^T$ respectively. After taking the Fourier transform we obtain the quantities
\begin{equation} \label{eq:2.9}
\mathcal{R}_n^{A_1:A_2}(q_1,q_2)=\frac{1}{(2\pi)^2}\int_{-\pi}^\pi\mathrm{d}\alpha\int_{-\pi}^\pi\mathrm{d}\beta\, e^{-i\alpha \left(q_1-\langle \hat{Q}_{A_1}\rangle\right)-i\beta \left(q_2+\langle \hat{Q}^T_{A_2}\rangle\right)}R_n\left(\alpha,\beta\right).
\end{equation}
The quantity $\mathcal{R}_1^{A_1:A_2}(q_1,q_2)$ may be interpreted as the joint probabilities of obtaining $q_1$ and $q_2$ for the measurements of $\hat{Q}_{A_1}$ and $-\hat{Q}_{A_2}^T$ respectively.
\section{Replica trick for charged moments} \label{sec3}
In this section, we briefly review the replica approach for the calculation of the R\'enyi negativity and the charged moments in 1+1 dimensional conformal field theory for two disjoint intervals.

We consider our subsystem to be $A=A_1\cup A_2$ as shown in figure \ref{fig:i}. In the replica method the reduced density matrix for the ground state is expressed by computing the euclidean path integral on the cut complex plane with boundary conditions along the cuts $(0^-,x)$ and $(0^+,x)$, where $x\in A_1\cup A_2$. We have
\begin{equation} \label{eq:3.3}
\rho\left(\{\phi_{A_1},\phi_{A_2}\},\{\phi_{A_1}',\phi_{A_2}'\}\right)=Z^{-1}\int\mathrm{D}\phi\prod_i\delta\left(\phi(0^-,x)-\phi_{A_i}\right)\delta\left(\phi(0^+,x)-\phi_{A_i}'\right)e^{-S_E},
\end{equation}
where $Z$ is the partition function on the complex plane and is introduced to normalise the trace. The trace of the integer powers of the reduced density matrix $\mathrm{Tr}\rho^n$ is obtained by evaluating the partition function $Z_n$ on the Riemann surface $\Sigma_n$. This Riemann surface is obtained by sewing together $n$ sheets along the cuts by setting $\phi(0^-,x)$ on the $i^{th}$ sheet equal to $\phi(0^+,x)$ on the $(i+1)^{th}$, where $i\in \{1,2,\cdots,n,n+1\equiv n\}$. We have
\begin{equation} \label{eq:3.4}
\mathrm{Tr}\rho^n=Z^{-n}\int\mathrm{D}\phi\,e^{-\int_{\Sigma_n}\mathrm{d}^2z\,\mathcal{L}_E[\phi]},
\end{equation}
where $\mathcal{L}_E$ is the corresponding euclidean lagrangian for the field $\phi$.
\begin{figure}
\centering 
\includegraphics[width=0.8\textwidth]{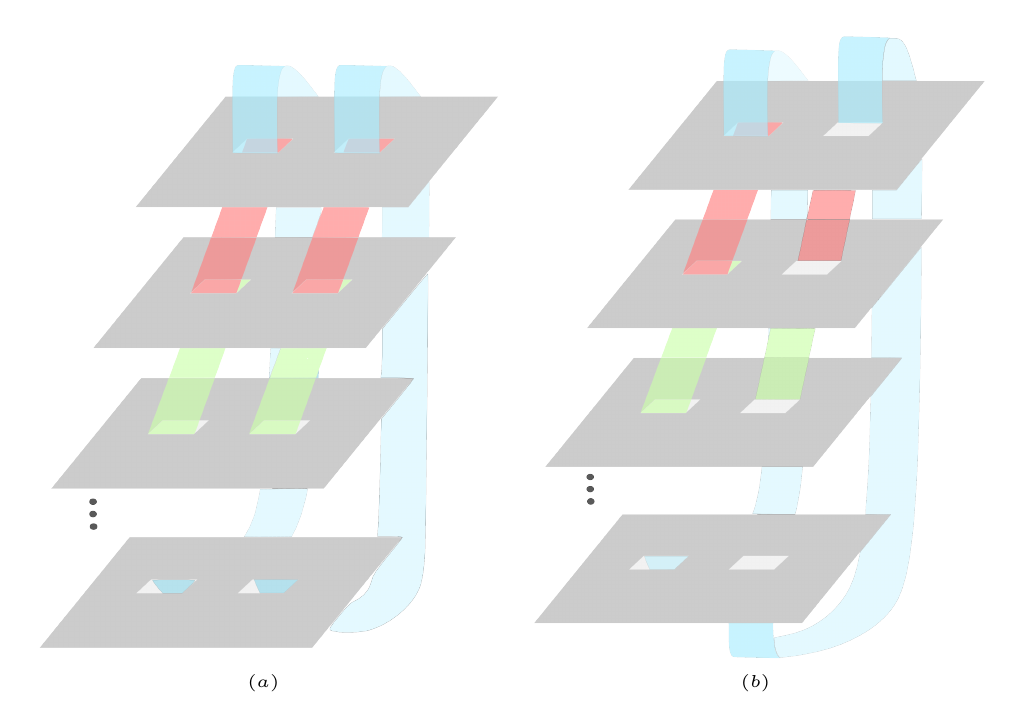}
\caption{\label{fig:ii} Riemann Surface $\Sigma_n$ obtained by sewing together $n$ copies of $(a)$ $\mathrm{\rho^n}$ $(b)$ $\mathrm{\left(\rho^{T_2}\right)^n}$ in evaluation of R\'enyi entropy and R\'enyi negativity respectively}
\end{figure}
The partition function on $\Sigma_n$ may also be evaluated from the model comprising $n$ copies of the field $\phi$, denoted by $\{\phi_i\}$, on the complex plane by introducing the twist fields $\mathcal{T}_n(x,y)$ and $\mathcal{\bar{T}}_n(x,y)$. Twist fields may be introduced corresponding to any global internal symmetry present in the theory. In the present case, we have a permutation symmetry among the $n$ fields, the twist fields introduced here correspond to the cyclic symmetry \cite{a39}
\begin{equation} \label{eq:3.6}
\begin{split}
\mathcal{T}_n\: :\: i\to i+1\: \mod n, \\
\mathcal{\bar{T}}_n\: :\: i\to i-1\: \mod n.
\end{split}
\end{equation}
Twist fields $\mathcal{T}_n(t,x)$ and $\mathcal{\bar{T}}_n(t,x)$ introduce the boundary conditions $\phi_i(t^-,y)=\phi_{i+1}(t^+,y)$, $\forall y\in(x,\infty)$ and $\phi_i(t^-,y)=\phi_{i-1}(t^+,y)$, $\forall y\in(t,\infty)$ respectively into the $n$ copy model path integral. In this framework the partition function on $\Sigma_n$ is given by \cite{a10}
\begin{equation} \label{eq:3.7}
Z_n=\left\langle\mathcal{T}_n(u_1)\mathcal{\bar{T}}_n(v_1)\mathcal{T}_n(u_2)\mathcal{\bar{T}}_n(v_2)\right\rangle.
\end{equation}
Here we have used the notation $u_i=(0,u_i)$ and $v_i=(0,v_i)$ for brevity. The twist fields are conformally invariant and have the scaling dimension
\begin{equation}
\Delta_n=\frac{c}{12}\left(\frac{1-n^2}{n}\right),
\end{equation}
where $c$ is the central charge of the CFT.

The partial transposition of $\rho$ is defined as
\begin{equation} \label{eq:3.2}
\langle{e^{(1)}_{i_1}e^{(2)}_{i_2}}|\rho^{T_2}|{e^{(1)}_{j_1}e^{(2)}_{j_2}}\rangle=\langle{e^{(1)}_{i_1}e^{(2)}_{j_2}}|\rho|{e^{(1)}_{j_1}e^{(2)}_{i_2}}\rangle,
\end{equation}
where $|{e^{(1)}_i}\rangle$ and $|{e^{(2)}_i}\rangle$ are the basis in $A_1$ and $A_2$ respectively.
In field theory, partial transposition is achieved by interchanging the upper and the lower cuts associated with $A_2$. As discussed in ref. \cite{a17} to compute $\mathrm{Tr}\left(\rho^{T_2}\right)^n$ we first reverse the cuts on $A_2$ and sew the $n$ sheets in a similar fashion as before, the resulting Riemann surface is shown in Figure \ref{fig:ii}. In the $n$ copy model the partition function on this Riemann surface is
\begin{equation} \label{eq:3.8}
R_n=\left\langle\mathcal{T}_n(u_1)\mathcal{\bar{T}}_n(v_1)\mathcal{\bar{T}}_n(u_2)\mathcal{T}_n(v_2)\right\rangle.
\end{equation}

We now discuss the replica method to compute the charged moments, and we will again restrict our discussion to the global $U(1)$ symmetry only. To find the symmetry resolution of entanglement measures in field theory one generally evaluates the charged moments introduced in section \ref{sre} and here we will focus on the charged moments for the R\'enyi negativity. In path integral evaluation, the presence of $e^{i\alpha\mathcal{\hat{Q}}}=e^{i\alpha \left(\hat{Q}_{A_1}-\hat{Q}_{A_2}\right)}$ (here we used $\hat{Q}_{A_2}^{T}=\hat{Q}_{A_2}$ in field theory) inside the trace changes the boundary conditions on the cuts between the different sheets. Since $e^{i\alpha \hat{Q}_{A_1}}$ and $e^{-i\alpha \hat{Q}_{A_2}}$ are the generators of $U(1)$ transformation in $A_1$ and $A_2$ respectively, they induce an additional phase factor in boundary conditions such that the total phase in $A_1$ and $A_2$ adds upto $\alpha$ and $-\alpha$ respectively. This phase has also been termed the Aharonov-Bohm flux. We now introduce the flux generating operators $\mathcal{V}_{\alpha}$; the operator $\mathcal{V}_{\alpha}$ introduces the phase boundary condition in the path integral \cite{a21}. For the present case, obtaining charged moments is equivalent to evaluating the correlation function of $\mathcal{V_\alpha}$ on the Riemann surface. We have the relation for the charged moments \cite{a22}
\begin{equation} \label{eq:3.9}
R_n(\alpha)\propto\left\langle \mathcal{V}_\alpha(u_1)\mathcal{V}_{-\alpha}(v_1)\mathcal{V}_{-\alpha}(u_2)\mathcal{V}_\alpha(v_2)\right\rangle_{\Sigma_n}R_n.
\end{equation}
The multi-charged moments are similarly obtained using the flux generating operators $\mathcal{V}_{\alpha}$, and as described in the next section are given by eq.(\ref{eq:4.13}).
\section{Charged moments for free compact boson} \label{section4}
In this section, we first introduce the theory of free compact boson and recapitulate the known results for R\'enyi entropy and R\'enyi negativity. We then proceed to calculate the multi-charged moments for negativity. We also numerically check our results against the tight-binding model.
\subsection{Free compact boson}
We consider the theory of massless boson in (1+1) dimension. The massless boson is a conformally invariant field with the central charge $c=1$ and its action given by
\begin{equation} \label{eq:4.1}
\mathcal{S}=\frac{1}{8\pi}\int\mathrm{d}^2x\,\partial_{\mu}\varphi\partial^{\mu}\varphi .
\end{equation}
The boson field $\varphi$ is compactified on the circle of radius R, i.e. we have the equivalence relation $\varphi\sim\varphi+2\pi kR$, where $k\in\mathbb{Z}$. Free compact boson is also the theory of the Luttinger liquid with the relation $R=\sqrt{\frac{2}{K}}$, where $K$ is the Luttinger parameter.

The $n=2$ R\'enyi entropy for the free compact boson in the case of two disjoint intervals was first studied in ref. \cite{a40} and was further generalised to integer values of $n$ in ref. \cite{a41}. The R\'enyi entropy in the latter work was obtained to be
\begin{equation} \label{eq:4.2}
\mathrm{Tr}\rho^n=c_n\left|\frac{(u_1-u_2)(v_1-v_2)}{(u_1-v_1)(u_1-v_2)(u_2-v_1)(u_2-v_2)}\right|^{2\Delta_n}\mathcal{F}_n(x),
\end{equation}
where $\Delta_n=\frac{1}{12}\left(\frac{1-n^2}{n}\right)$ is the scaling dimension of the twist operator $\mathcal{T}_n$ introduced in section \ref{sec3}, $c_n$ is a non-universal constant, and $\mathcal{F}_n(x)$ is a function of the cross ratio $x$
\begin{equation*}
x=\frac{(v_1-u_1)(v_2-u_2)}{(v_2-v_1)(u_2-u_1)}.
\end{equation*}
This definition also implies that $x\in(0,1)$. The function $\mathcal{F}_n(x)$ depends on the local conformal operator content as well and is given by
\begin{equation} \label{eq:4.3}
\mathcal{F}_n(x)=\frac{\Theta\left(0|K\Gamma(x)\right)\Theta\left(0|\Gamma(x)/K\right)}{\Theta\left(0|\Gamma(x)\right)^2},
\end{equation}
where $\Theta$ is the Riemann Siegel theta function. It is given by
\begin{equation} \label{eq:4.4}
\Theta\left[\begin{array}{l}
\varepsilon \\
\delta
\end{array}\right](\boldsymbol{u} \mid \Omega) \equiv \sum_{\boldsymbol{m} \in \mathbb{Z}^{n-1}} e^{i \pi(\boldsymbol{m}+\boldsymbol{\varepsilon})^t \cdot \Omega \cdot (\boldsymbol{m}+\boldsymbol{\varepsilon})+2 \pi i(\boldsymbol{m}+\boldsymbol{\varepsilon})^t \cdot(\boldsymbol{u}+\delta)},
\end{equation}
where $\Omega$ must be a $(n-1)\times (n-1)$ symmetric matrix with a positive definite imaginary part, while the characteristics $\varepsilon$, $\delta \in\left(\mathbb{Z}/2\right)^{n-1}$ and $\boldsymbol{u}\in \mathbb{C}^{n-1}$. The quantity $\Gamma(x)$ in equation (\ref{eq:4.3}) is the period matrix of the Riemann surface and is given by
\begin{equation} \label{eq:4.5}
\begin{split}
&\Gamma_{rs}(x)=\frac{i2}{n}\sum_{k=1}^{n-1}\sin\left(\frac{\pi k}{n}\right)\beta_{k/n}(x)\cos\left[\frac{2\pi k}{n}(r-s)\right],\\ 
&\text{with}\qquad \beta_{k/n}(x)=\frac{F_{k/n}(1-x)}{F_{k/n}(x)},
\end{split}
\end{equation}
where $F_{k/n}(x)\equiv {}_{2}F_{1}\left(k/n,1-k/n;1;x\right)$ is the hypergeometric function. The function $\mathcal{F}(x)$ is invariant under the inversion $K\to1/K$ and under the self dual limit $K\to 1$ it becomes unity. In the limit $x\to 0$, we have the case of infinite separation and the function $\mathcal{F}(x)$ is unity. For the $x\to 1$ limit, we reach the case of a single interval and the function $\mathcal{F}(x)$ again becomes unity, as it should. The case for multiple disjoint intervals was considered in ref. \cite{a42}.

The R\'enyi negativity for the same case was studied in ref. \cite{a16,a17}. Analytic expression for the R\'enyi negativity was obtained by studying the correlation $\left\langle\mathcal{T}_n(u_1)\mathcal{\bar{T}}_n(v_1)\mathcal{\bar{T}}_n(u_2)\mathcal{T}_n(v_2)\right\rangle$. It is given by
\begin{equation} \label{eq:4.6}
R_n=c_n\left|\frac{(u_1-v_2)(v_1-u_2)}{(v_1-u_1)(u_2-u_1)(v_2-v_1)(v_2-u_2)}\right|^{2\Delta_n}\mathcal{F}_n(x).
\end{equation}
In this case the cross ratio $x$ is
\begin{equation} \label{eq:4.7}
x=\frac{(v_1-u_1)(u_2-v_2)}{(u_2-v_1)(v_2-u_1)},
\end{equation}
and from this definition we have $x\in (-\infty,0)$. The function $\mathcal{F}_n(x)$ in this range is given by
\begin{equation} \label{eq:4.8}
\mathcal{F}_n(x)=\frac{\Theta\left(0|T(x)\right)}{\prod_{k=1}^n F_{k/n}(x)},
\end{equation}
where $T(x)$ is a $2(n-1)\times2(n-1)$ matrix. $T(x)$ may be written in terms of two real $(n-1)\times(n-1)$ matrices $\mathcal{R}$ and $\mathcal{I}$
\begin{equation} \label{eq:4.9}
T(x)=\begin{pmatrix}
iK\mathcal{I} & \mathcal{R}\\
\mathcal{R} & i\mathcal{I}/K
\end{pmatrix}.
\end{equation}
The Riemann period matrix $\tau$ for the Riemann surface $\Sigma_n(x)$ has the relation $\tau=\mathcal{R}+i\mathcal{I}$, where the period matrix is given by
\begin{equation} \label{eq:4.10}
\begin{split}
&\tau_{rs}=\frac{2}{n}\sum_{k=1}^{n-1}\sin\left(\frac{\pi k}{n}\right)\left(\alpha_{k/n}(x)+i\beta_{k/n}(x)\right)e^{i\left(\frac{2\pi k}{n}(r-s)\right)},\\
&\text{with}\qquad\alpha_{k/n}+i\beta_{k/n}(x)=i\frac{F_{k/n}(1-x)}{F_{k/n}(x)}.
\end{split}
\end{equation}
In Appendix \ref{A1}, we also obtain the period matrix for the case of R\'enyi negativity by studying the normalised holomorphic differential on a homology basis for the Riemann surface $\Sigma_n(x)$. The function $\mathcal{F}_n(x)$ is invariant under the inversion of $K$ in this range of $x$ as well however, it is not unity in the self dual limit. In the limit $x\to 0$, we have the case of infinite separation and $\mathcal{F}_n(x)$ becomes unity. 
\subsection{Multi charged moments for negativity}
The free compact boson has a global $U(1)$ symmetry due to the invariance under the field transformations $\varphi\to\varphi+2\pi mR,\,m\in\mathbb{Z}$. The conserved charge associated with this symmetry is $\hat{Q}=\frac{1}{2\pi}\int\mathrm{d}x\,\partial_x\varphi$. The corresponding flux generator $\mathcal{V}_\alpha(z)$ is the vertex operator \cite{a21}
\begin{equation} \label{eq:4.11}
\mathcal{V}_\alpha(z)=e^{i\frac{\alpha\varphi(z)}{2\pi}},
\end{equation}
with the conformal weight $h_\alpha^{\mathcal{V}}$ given by
\begin{equation} \label{eq:4.12}
h_\alpha^{\mathcal{V}}=\left(\frac{\alpha}{2\pi}\right)^2\frac{K}{2}.
\end{equation}
To obtain the charged moments, the flux generators are placed on the branch points of the Riemann surface, and so the multi-charged moments for the R\'enyi negativity are
\begin{equation} \label{eq:4.13}
R_n(\alpha,\beta)\propto\left\langle \mathcal{V}_\alpha(u_1)\mathcal{V}_{-\alpha}(v_1)\mathcal{V}_{-\beta}(u_2)\mathcal{V}_\beta(v_2)\right\rangle_{\Sigma_n}R_n.
\end{equation}
The multi-charged moments for the same setting in the case of the R\'enyi entropies were recently studied in ref. \cite{a30}. They also used these results to obtain the charge resolved R\'enyi entropies and mutual information. The charged moments for the R\'enyi negativity for the case of complex boson were obtained in ref. \cite{a33}. However, we found an error in these calculations and so we dedicate appendix \ref{A5} for a very brief discussion on the complex boson since this problem is similar to the present one. To simplify the calculations, we use the global conformal invariance to map the points $u_1\to 0$, $v_1\to x$, $u_2\to \infty$ and $v_2\to 1$ via eq.(\ref{eq:A.1}) \cite{a43}, where $x$ is given by eq.(\ref{eq:4.7}). Under this transformation, we have the Riemann surface $\Sigma_n(x)$ generated by the curve
\begin{equation} \label{eq:4.14}
y^n=\frac{z(z-1)}{z-x}.
\end{equation}
The correlation function $\left\langle \mathcal{V}_{-\alpha}(x)\mathcal{V}_\alpha(0)\mathcal{V}_\beta(1)\mathcal{V}_{-\beta}(\infty)\right\rangle_{\Sigma_n}$ is just the correlation function of the vertex operators on the Riemann surface $\Sigma_n(x)$. These correlation have been studied in ref. \cite{a44,a45}. The four-point correlation function is of the following form
\begin{equation} \label{eq:4.15}
\begin{split}
&\left\langle \mathcal{V}_{\alpha_1}(z_1)\mathcal{V}_{\alpha_2}(z_2)\mathcal{V}_{\alpha_3}(z_3)\mathcal{V}_{\alpha_4}(z_4)\right\rangle_{\Sigma_n}=\\
&\hspace{0.8in}\prod_{1\leq i<i'\leq 4}\left| E(z_i,z_{i'})e^{-\pi \mathrm{Im}|\boldsymbol{w}(z_i)-\boldsymbol{w}(z_{i'})|^t\cdot\mathrm{Im}|\tau(x)^{-1}|\cdot\mathrm{Im}|\boldsymbol{w}(z_i)-\boldsymbol{w}(z_{i'})|}\right|^{\alpha_i\alpha_{i'}K/2\pi^2},
\end{split}
\end{equation}
with the condition $\sum_i \alpha_i=0$. In our case we have $z_1=x$, $z_2=0$, $z_3=1$, and $z_4=\infty$ with corresponding $\alpha_1=-\alpha$, $\alpha_2=\alpha$, $\alpha_3=\beta$ and $\alpha_4=-\beta$. The map $\boldsymbol{w}(z)=\left(w_1(z),\cdots,w_{n-1}(z)\right)$ is the Abel-Jacobi map and is discussed later in this section. The quantity $E(z_i,z_{i'})$ is the prime form of the Riemann surface $\Sigma_n(x)$ \cite{a46,a47} and is given by
\begin{equation} \label{eq:4.16}
E(z_i,z_{i'})=\frac{\Theta_{\boldsymbol{\Delta}}\left(\boldsymbol{w}(z_i)-\boldsymbol{w}(z_{i'})|\tau(x)\right)}{h_{\boldsymbol{\Delta}}(z_i)h_{\boldsymbol{\Delta}}(z_{i'})},
\end{equation}
where ${\boldsymbol{\Delta}}=({\boldsymbol{\epsilon}},{\boldsymbol{\delta}})$ is a non-singular odd half characteristics. The prime form $E(z_i,z_{i'})$ is independent of the choice of $\boldsymbol{\Delta}$. In this work, we will use $\boldsymbol{\epsilon}=\boldsymbol{\delta}=
(1/2,0,\cdots,0)$ and use the shorthand  $\boldsymbol{\Delta}=\mathbf{\frac{1}{2}}$ to denote this. The quantity $h_{\boldsymbol{\Delta}}(z_i)$ is a holomorphic 1-form and is given in terms of the holomorphic normalised differentials $\nu_r(z)$ (given by eq.(\ref{eq:A.15}))
\begin{equation} \label{eq:4.17}
h_{\boldsymbol{\Delta}}(z_i)=\left(\sum_{r=1}^{n-1}\nu_r(z_i)\partial_{u_r}\Theta_{\boldsymbol{\Delta}}(\boldsymbol{u}|\tau)|_{u=0}\right)^{\frac{1}{2}}.
\end{equation}
However, the normalised holomorphic differentials $\nu_r$ at the points $x$, $0$, $1$, and $\infty$ (i.e. the branch points of the Riemann surface $\Sigma_n(x)$) are singular with the leading order singular behaviour $\nu_r(z+\epsilon)\sim \epsilon^{\frac{1-n}{n}}\nu^{(*)}_r(z)+O(\epsilon^{1/n})$ near the branch points, where
\begin{equation} \label{eq:4.18}
\nu^{(*)}_r(z)=\left\{
\begin{array}{ll}
\left(x(1-x)\right)^{-1/n}e^{\frac{-i\pi(4r-3)}{n}}Q_{r,n}(x), & z=x,\\
-x^{-1/n}Q_{r,n}(x), & z=0,\\
(1-x)^{-1/n}Q_{r,n}(x), & z=1,\\
e^{\frac{-i4\pi(r-1)}{n}}Q_{r,n}(x), & z=\infty,
\end{array}
\right.
\end{equation}
with $Q_{r,n}(x)=e^{\frac{i2\pi(r-1)}{n}}\frac{\sin{\pi/n}}{n\pi F_{1/n}(x)}$. Consequently $h_{\boldsymbol{\Delta}}$ is also singular near the branch points and hence the generic correlation function in eq.\eqref{eq:4.15} is not well defined at the branch points. To resolve this issue, the regularised vertex operators $\mathcal{V}^{(*)}(z_i)$ were introduced to remove the leading order singularities in ref. \cite{a30}. Following this reference we introduce
\begin{equation} \label{eq:4.19}
\mathcal{V}_\alpha^{(*)}(z)=\lim_{\epsilon\to 0}\left(\kappa_n \epsilon^{\frac{1-n}{n}}\right)^{2h_\alpha^\nu}\mathcal{V}_\alpha(z+\epsilon),
\end{equation}
where $\kappa_n$ is a surface dependent global rescaling factor and will be fixed later in this section. The generic four point correlation of the regularised vertex operators on the Riemann surface $\Sigma_n(x)$ is given by
\begin{equation} \label{eq:4.20}
\begin{split}
&\left\langle \mathcal{V}^{(*)}_{-\alpha}(x)\mathcal{V}^{(*)}_{\alpha}(0)\mathcal{V}^{(*)}_{\beta}(1)\mathcal{V}^{(*)}_{-\beta}(\infty)\right\rangle_{\Sigma_n}=\\
&\hspace{0.01in}(\kappa_n)^{(\alpha^2+\beta^2)K/2\pi^2}\prod_{1\leq i<i'\leq 4}\left| E^{(*)}(z_i,z_{i'})e^{-\pi \mathrm{Im}|\boldsymbol{w}(z_i)-\boldsymbol{w}(z_{i'})|^t\cdot\mathrm{Im}|\tau(x)^{-1}|\cdot\mathrm{Im}|\boldsymbol{w}(z_i)-\boldsymbol{w}(z_{i'})|}\right|^{\alpha_i\alpha_{i'}K/2\pi^2},
\end{split}
\end{equation}
where the regularised prime form $E^{(*)}(z_i,z_{i'})$ for the branch points is now defined in terms of $h^{(*)}_{\mathbf{\frac{1}{2}}}(z_i)$,
\begin{equation} \label{eq:4.21}
h^{(*)}_{\mathbf{\frac{1}{2}}}(z_i)=\left(\sum_{r=1}^{n-1}\nu^*_r(z_i)\partial_{u_r}\Theta_{\mathbf{\frac{1}{2}}}(\boldsymbol{u}|\tau)|_{u=0}\right)^{\frac{1}{2}}.
\end{equation} 
The regularised prime forms were conjectured and numerically verified in ref. \cite{a30} to be simple algebraic functions of $x$ for $x\in(0,1)$. We extend and numerically verify (in Appendix \ref{A2}) these conjectures for our case to be
\begin{align}
|E^{(*)}(x,0)|&=n|x|^{1/n}, \label{eq:4.22}\\
|E^{(*)}(1,\infty)|&=n, \label{eq:4.23} \\
|p(x,0,1,\infty)|&=|1-x|^{1/n}, \label{eq:4.24}
\end{align} 
where $p(x,0,1,\infty)$ is a cross ratio function on the Riemann surface and is given by
\begin{equation} \label{eq:4.24i}
p(x,0,1,\infty)=\frac{E(x,1)E(0,\infty)}{E(x,\infty)E(0,1)}=\frac{E^{(*)}(x,1)E^{(*)}(0,\infty)}{E^{(*)}(x,\infty)E^{(*)}(0,1)}.
\end{equation}
The Abel-Jacobi map $\boldsymbol{w}(z)$ is defined from the Riemann surface $\Sigma_n(x)$ to the quotient space $\mathbb{C}^{n-1}/\Lambda$ (which is a $n-1$ genus torus in the present case), where $\Lambda=\mathbb{Z}^{n-1}+\tau(x)\mathbb{Z}^{n-1}$. In terms of the normalised holomorphic differentials $\nu_r(z)$ the components of $\boldsymbol{w}(z)$ are
\begin{equation} \label{eq:4.25}
w_r(z)=\int_0^z \mathrm{d}z'\nu_r(z')\; \mod{\Lambda}.
\end{equation}
Using the expression for the normalised holomorphic differentials $\nu_r(z)$ in eq.(\ref{eq:A.15}) $\boldsymbol{w}$ is evaluated at points $x$, $0$, $1$, and $\infty$ to be 
\begin{align}
\boldsymbol{w}(x) &=\boldsymbol{q}, \label{eq:4.26} \\
\boldsymbol{w}(0) &=\mathbf{0}, \label{eq:4.27}\\
\boldsymbol{w}(1) &=\boldsymbol{q}+\boldsymbol{\tilde{p}}(x)+i\boldsymbol{p}(x),\label{eq:4.28} \\
\boldsymbol{w}(\infty) &=\boldsymbol{\tilde{p}}(x)+i\boldsymbol{p}(x).\label{eq:4.29}
\end{align}
The components of $\boldsymbol{q}$, $\boldsymbol{\tilde{p}}(x)$ and $\boldsymbol{p}(x)$ are given by
\begin{align}
q_r&=\frac{1}{n}, \label{eq:4.30}\\
\tilde{p}_r&=-\frac{1}{n}\sum_{l=1}^{n-1}\left[\cos\left[\frac{2\pi l(r-1)}{n}\right]\sin\left[\frac{\pi l}{n}\right]+\sin\left[\frac{2\pi l(r-1)}{n}\right]\cos\left[\frac{\pi l}{n}\right]\right]\alpha_{l/n}(x),\label{eq:4.31}\\
{p}_r&=-\frac{1}{n}\sum_{l=1}^{n-1}\left[\cos\left[\frac{2\pi l(r-1)}{n}\right]\sin\left[\frac{\pi l}{n}\right]+\sin\left[\frac{2\pi l(r-1)}{n}\right]\cos\left[\frac{\pi l}{n}\right]\right]\beta_{l/n}(x),\label{eq:4.32}
\end{align}
here we used the equivalence $\left(-1+1/n,1/n,\cdots,1/n\right)\sim\left(1/n,1/n,\cdots,1/n\right)$ under $\mod{\Lambda}$ in evaluation of $\boldsymbol{q}$. Using eq.\eqref{eq:4.22}-eq.\eqref{eq:4.24} and eq.\eqref{eq:4.26}-eq.(\eqref{eq:4.29} in eq.\eqref{eq:4.20}, the correlation function of the regularised vertex operators simplifies to
\begin{equation} \label{eq:4.33}
\left\langle \mathcal{V}^{(*)}_{-\alpha}(x)\mathcal{V}^{(*)}_{\alpha}(0)\mathcal{V}^{(*)}_{\beta}(1)\mathcal{V}^{(*)}_{-\beta}(\infty)\right\rangle_{\Sigma_n}=\left(\frac{\kappa_n}{n}\right)^{(\alpha^2+\beta^2)K/2\pi^2}|x|^{-{\alpha^2 K}/{2\pi^2 n}}|1-x|^{-{\alpha\beta}/{2\pi^2 n}}.
\end{equation}

Taking the global conformal transformation $0\to u_1$, $x\to v_1$, $1\to v_2$, and $\infty\to u_2$, the multi-charged moments $R_n^{A_1:A_2}(\alpha,\beta)$ in eq.\eqref{eq:4.13} with the vertex operators replaced with the regularised vertex operators become
\begin{equation} \label{eq:4.34}
R_n^{A_1:A_2}(\alpha,\beta)= c_{n,\alpha,\beta} \left(\frac{\kappa_n}{n}\right)^{\frac{(\alpha^2+\beta^2)K}{2\pi^2 n}}\ell_1^{-\frac{\alpha^2 K}{2\pi^2n}}\ell_2^{-\frac{\beta^2 K}{2\pi^2n}}{|1-x|}^{-\frac{\alpha\beta K}{2\pi^2 n}}(\ell_1\ell_2|1-x|)^{2\Delta_n}\mathcal{F}_n(x),
\end{equation}
where $\ell_1=|v_1-u_1|$ and $\ell_2=|v_2-u_2|$. To fix $\kappa_n$ we consider the limit $d\to\infty$, where $d=|u_2-v_1|$. In this limit $x\to0$, consequently the function $\mathcal{F}_n(x)\to 1$ and we would expect the decomposition $R_n^{A_1:A_2}(\alpha,\beta)=Z_n^{A_1}(\alpha)Z_n^{A_2}(\beta)$. This is achieved by setting $\kappa_n=n$. This decomposition also implies that the non-universal constant $c_{n,\alpha,\beta}$ is factorisable into $\alpha$ and $\beta$ terms. In this work, we will assume that the non-universal constant $c_{n,\alpha,\beta}$ is well approximated to the leading order in $\alpha$ and $\beta$ by $c_{n,\alpha,\beta}\sim c_n\lambda_n^{-(\alpha^2+\beta^2) K/2\pi^2n}$ \cite{a30}. Using this approximation we have 
\begin{align}
R_n^{A_1:A_2}(\alpha,\beta)&= \tilde{\ell}_1^{-\frac{\alpha^2 K}{2\pi^2n}}\tilde{\ell}_2^{-\frac{\beta^2 K}{2\pi^2n}}{|1-x|}^{-\frac{\alpha\beta K}{2\pi^2n}}R_n, \label{eq:4.35}\\
R_n(\alpha)&= \left(\tilde{\ell}_1\tilde{\ell}_2|1-x|\right)^{-\frac{\alpha^2 K}{2\pi^2 n}}R_n, \label{eq:4.36}
\end{align}
here $\tilde{\ell_i}=\lambda_n \ell_i$. For the case of adjacent intervals we take the limit $d\to 0$. In this limit
\begin{equation} \label{eq:4.37}
|1-x|\sim \lim_{d\to 0}\frac{\ell_1 \ell_2}{d(\ell_1+\ell_2)},
\end{equation}
\begin{figure}
\centering 
\includegraphics[width=1\textwidth]{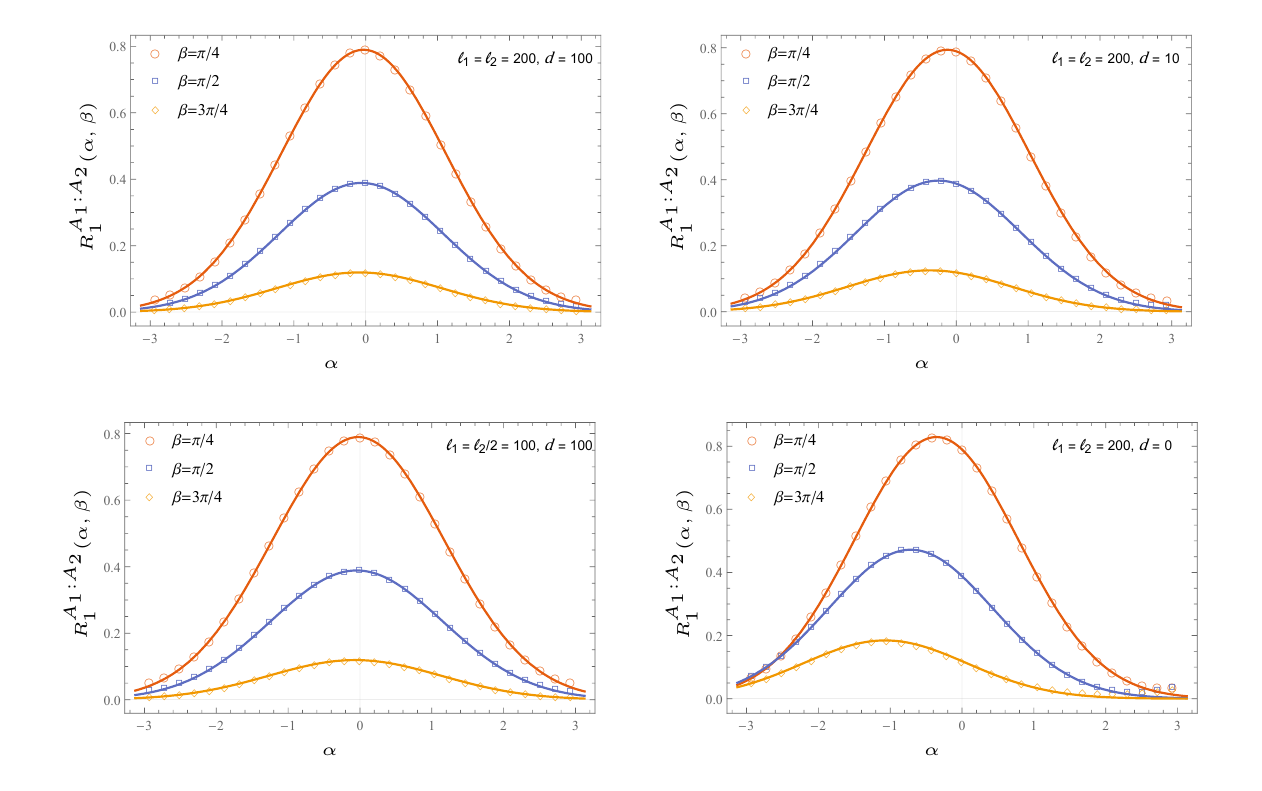}
\caption{\label{fig:iii} Plots for the multi-charged moments. The continuous lines are plots of $R_1^{A_1:A_2}(\alpha,\beta)$ (given by using $n=1$ in eq.\eqref{eq:4.35} for the top two and bottom left plots and in eq.\eqref{eq:4.38} for the bottom right plot) for $K=1$ as a function of $\alpha$ at different values of $\beta$. The discrete points are the plots of the numerically evaluated multi-charged moments for the tight-binding model given in eq.\eqref{eq:C.5}. In the top plots, $\ell_1=\ell_2$ and $d$ is varied. In bottom left plot, $\ell_1\neq\ell_2$ and the bottom right plot is for the adjacent interval case i.e. $d=0$.} 
\end{figure}
however $d$ must be absorbed into the UV cut-off. In this case, we have $v_1=u_2$ and the two vertex operator coincide. We can approximate the non-universal constant to the leading order in $\alpha$ and $\beta$ as $c_{n,\alpha,\beta}\sim c_{n}\lambda_n^{-(\alpha^2+\beta^2+\alpha\beta)K/2\pi^2 n}$ and hence the charged moments for the case of adjacent intervals are 
\begin{align}
R_n^{A_1:A_2}(\alpha,\beta)&=\tilde{\ell}_1^{-\frac{\alpha^2 K}{2\pi^2 n}-\frac{\alpha\beta K}{2\pi^2 n}}\tilde{\ell}_2^{-\frac{\beta^2 K}{2\pi^2 n}-\frac{\alpha\beta K}{2\pi^2 n}}{(\tilde{\ell}_1+\tilde{\ell}_2)}^{\frac{\alpha\beta K}{2\pi^2 n}}R_n, \label{eq:4.38}\\
R_n(\alpha)&=\left(\tilde{\ell}_1\tilde{\ell}_2\right)^{-\frac{\alpha^2 K}{\pi^2 n}}{(\tilde{\ell}_1+\tilde{\ell}_2)}^{\frac{\alpha^2 K}{2\pi^2 n}}R_n. \label{eq:4.39}
\end{align}
We note that eq.\eqref{eq:4.39} matches the result of ref. \cite{a22}.
\begin{figure}
\centering 
\includegraphics[width=1\textwidth]{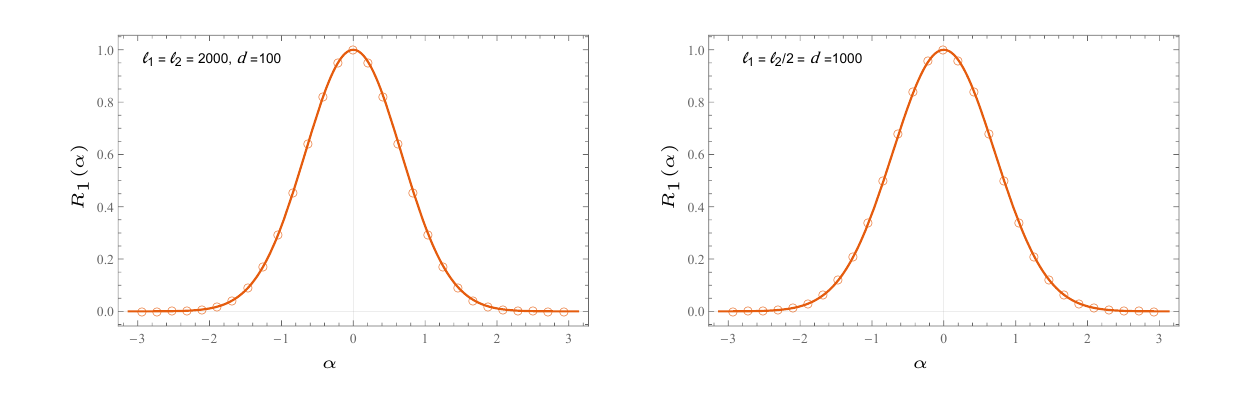}
\caption{\label{fig:iv} Plots for charged moments $R_1(\alpha)$. The continuous lines are plots of $R_1(\alpha)$ (given by using $n=1$ in eq.\eqref{eq:4.36}) for $K=1$ as a function of $\alpha$. The discrete points are the plots of numerically evaluated charged moments for the tight-binding model given by using $\beta=\alpha$ in eq.\eqref{eq:C.5}}
\end{figure}
\begin{figure}
\centering 
\includegraphics[width=1\textwidth]{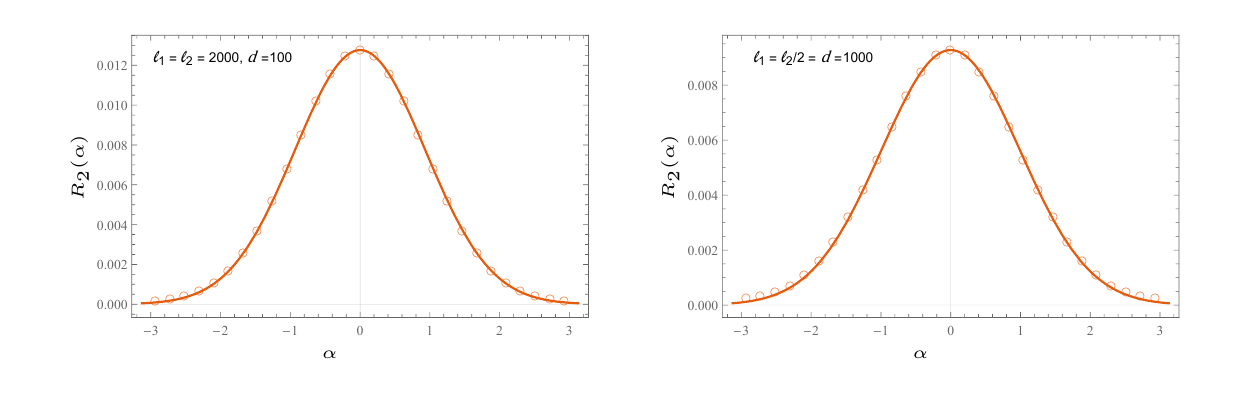}
\caption{\label{fig:v} Plots for charged moments $R_2(\alpha)$. The continuous lines are plots of $R_2(\alpha)$ (given by using $n=2$ in eq.\eqref{eq:4.36}) for $K=1$ as a function of $\alpha$. The discrete points are the plots of numerically evaluated charged moments for the tight-binding model given in eq.\eqref{eq:C.6}}
\end{figure}
As discussed in ref. \cite{a19i,a48} the R\'enyi entropy and the R\'enyi negativity for the two interval case in free fermions and the free compact boson at self dual radius match. In the figures \ref{fig:iii}, \ref{fig:iv} and \ref{fig:v}, we have matched our analytical results for $R_1^{A_1:A_2}(\alpha,\beta)$, $R_1(\alpha)$, and $R_2(\alpha)$ against the numerical results of the tight binding model (Appendix \ref{A4}). We used $\lambda_1=9.682$ and $\lambda_2=12.402$ \cite{a23}, using these parameters we have a very good match with the numerical results.
\section{Charge imbalance resolved R\'enyi negativity} \label{section5}
In this section, we take the Fourier transforms of the charged moments calculated in section \ref{section4} and thus obtain the charge imbalance resolved R\'enyi negativity. We also verify these results against the tight binding model.

First we take the Fourier transform of the multi charged moments. To take the Fourier transform  of eq.\eqref{eq:4.35} we take the Gaussian approximation in both the integration variables and using the standard techniques for solving the multi dimensional Gaussian integral. We have
\begin{equation} \label{eq:5.1}
\begin{split}
&\mathcal{R}_n^{A_1:A_2}(q_1,q_2)=\\
&\hspace{0.4in}\pi nR_n\frac{\exp\left\{-\frac{2\pi^2 n}{K}\frac{\left(q_1-\langle\hat{Q}_1\rangle\right)^2\ln{(\tilde{\ell}_2)}+\left(q_2+\langle\hat{Q}^T_2\rangle\right)^2\ln{(\tilde{\ell}_1)}+\left(q_1-\langle\hat{Q}_1\rangle\right) \left(q_2+\langle\hat{Q}^T_2\rangle\right)\ln{|1-x|}}{4\ln{(\tilde{\ell}_1)}\ln{(\tilde{\ell}_2)}-\ln{\left(|1-x|\right)^2}}\right\}}{K\left(4\ln{(\tilde{\ell}_1)}\ln{(\tilde{\ell}_2)}-\ln{\left(|1-x|\right)^2}\right)^{1/2}} 
\end{split}
\end{equation}
we observe that the quantity $\mathcal{R}_n^{A_1:A_2}(q_1,q_2)$ is a two dimensional Gaussian distribution in variables $q_1$ and $q_2$. Similarly, the result for the adjacent interval is obtained from the Fourier transform of eq.\eqref{eq:4.38}
\begin{equation}\label{eq:5.2}
\begin{split}
&\mathcal{R}_n^{A_1:A_2}(q_1,q_2)=\\
&\hspace{0.1in}\pi nR_n \frac{\exp\left\{-\frac{2\pi^2 n}{K}\frac{\left(q_1-\langle\hat{Q}_1\rangle\right)^2\ln{(\tilde{\ell}_2)}+\left(q_2+\langle\hat{Q}^T_2\rangle\right)^2\ln{(\tilde{\ell}_1)}+\left(q_1-\langle\hat{Q}_1\rangle\right) \left(q_2+\langle\hat{Q}^T_2\rangle\right)\ln{\tilde{\ell}_1\tilde{\ell}_2/(\tilde{\ell}_1+\tilde{\ell}_2}}{4\ln{(\tilde{\ell}_1)}\ln{(\tilde{\ell}_2)}-\ln{\left(\tilde{\ell}_1\tilde{\ell}_2/(\tilde{\ell}_1+\tilde{\ell}_2\right)^2}}\right\}}{K\left(4\ln{(\tilde{\ell}_1)}\ln{(\tilde{\ell}_2)}-\ln{\left(\tilde{\ell}_1\tilde{\ell}_2/(\tilde{\ell}_1+\tilde{\ell}_2)\right)^2}\right)^{1/2}}.
\end{split}
\end{equation}
In the infinite separation limit i.e. $|1-x|\to1$, the Gaussian distribution above factors into two independent Gaussian distributions indicating that measurements on $A_1$ and $A_2$ become independent of each other
\begin{equation} \label{eq:5.3}
\begin{split}
&\mathcal{R}_n^{A_1:A_2}(q_1,q_2)=\\
&R_n\sqrt{\frac{\pi n}{2K\ln{(\tilde{\ell}_1)}}}\exp\left\{-\frac{\pi^2 n \left(q_1-\langle\hat{Q}_1\rangle\right)^2}{2K\ln{(\tilde{\ell}_1)}}\right\}\sqrt{\frac{\pi n}{2K\ln{(\tilde{\ell}_2)}}}\exp\left\{-\frac{\pi^2 n \left(q_2+\langle\hat{Q}^T_2\rangle\right)^2}{2K\ln{(\tilde{\ell}_2)}}\right\}.
\end{split}
\end{equation}
\begin{figure}
\centering 
\includegraphics[width=1\textwidth]{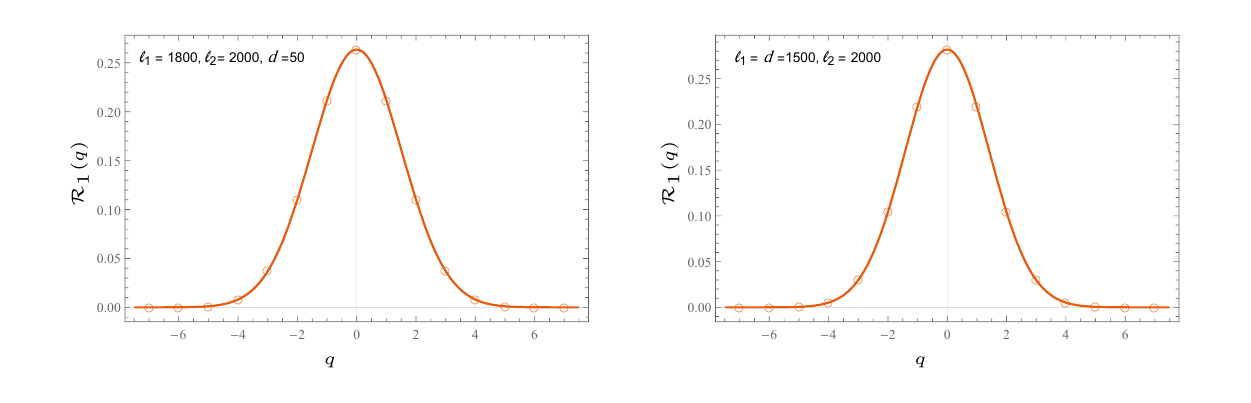}
\caption{\label{fig:vi} Plots for charged imbalance resolved R\'enyi negativity for $n=1$. The continuous lines are plots of $\mathcal{R}_1(q)$ from eq.\eqref{eq:5.4}) with $K=1$ as a function of charge $q$. The discrete points are the plots of numerically evaluated charged imbalance resolved R\'enyi negativities for tight-binding model given by plugging eq.\eqref{eq:C.5}, with $\beta=\alpha$, in eq.\eqref{eq:C.7}.}
\end{figure}
\begin{figure}
\centering 
\includegraphics[width=1\textwidth]{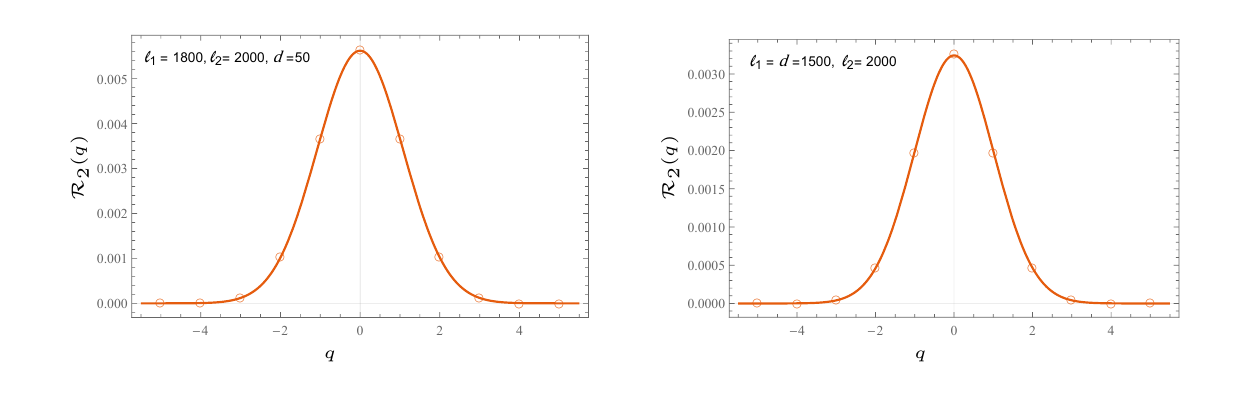}
\caption{\label{fig:vii} Plots for charged imbalance resolved R\'enyi negativity for $n=2$. The continuous lines are plots of $\mathcal{R}_2(q)$ from eq.\eqref{eq:5.4} with $K=1$ as a function of charge $q$. The discrete points are the plots of numerically evaluated charged imbalance resolved R\'enyi negativities for tight-binding model given by plugging eq.\eqref{eq:C.6} in eq.\eqref{eq:C.7}.}
\end{figure}

The charge imbalance resolved negativity are obtained by taking the Fourier transform of eq.\eqref{eq:4.36}. Again using the Gaussian approximation, we have the following result
\begin{equation} \label{eq:5.4}
\mathcal{R}_n(q)=\frac{\sqrt{\pi n}R_n}{\sqrt{2K\ln{\left(\tilde{\ell_1}\tilde{\ell_1}|1-x|\right)}}}\exp\left\{-\frac{\pi^2 n\left(q-\langle\mathcal{\hat{Q}}\rangle\right)^2}{2Kln{\left(\tilde{\ell_1}\tilde{\ell_1}|1-x|\right)}}\right\}.
\end{equation}
We note that the charge imbalance resolved negativity has a Gaussian distribution in $q$ about the mean $<\hat{\mathcal{Q}}>$. The Luttinger parameter appears as an overall factor in the exponent, and the R\'enyi negativity appears as an overall factor similar to the case of charged resolved R\'enyi entropy. 
In the case of adjacent interval, we take the Fourier transform of eq.\eqref{eq:4.39}
\begin{equation} \label{eq:5.5}
\mathcal{R}_n(q)=\frac{\sqrt{\pi n}R_n}{\sqrt{2K\ln{\left(\tilde{\ell_1}^2\tilde{\ell_2}^2/(\tilde{\ell_1}+\tilde{\ell_2})\right)}}}\exp\left\{-\frac{\pi^2 n\left(q-\langle\mathcal{\hat{Q}}\rangle\right)^2}{2Kln{\left(\tilde{\ell_1}^2\tilde{\ell_2}^2/(\tilde{\ell_1}+\tilde{\ell_2})\right)}}\right\}.
\end{equation}
In figures \ref{fig:vi} and \ref{fig:vii}, we have matched our analytic results for the disjoint interval case  against the numerical result of tight binding model. As done in section \ref{section4}, we considered $n=1$ and $n=2$ case for the numerical test.
\section{Conclusion} \label{section6}
In this work, we obtained the symmetry decomposition of the R\'enyi negativity into the charge imbalance sectors in the ground state of the free compact boson of arbitrary compactification radius for the case of two disjoint intervals.

The free compact boson is the conformal field theory of the Luttinger liquids and has a global $U(1)$ symmetry. We determined the charged moments for the R\'enyi negativity by evaluating the four-point correlation function of the flux generating vertex operators corresponding to the $U(1)$ symmetry on the Riemann surface $\Sigma_n(x)$  obtained from replica method. The charged moments were reduced to simple algebraic functions of interval lengths and distances by extending the conjectures for Prime forms in ref. \cite{a30} to the present case. The charge imbalance resolved R\'enyi negativity was obtained by taking the Fourier transform of the charged moments. We found that the R\'enyi negativity assumes a Gaussian form in the charge imbalance sectors. We also numerically checked our results against the tight-binding model or the free fermions on a lattice, and we indeed found a very good match between the analytic and the numerical results. We also briefly discussed a correction to the charged moments for the case of compact complex boson with global $U(1)$ symmetry obtained in ref. \cite{a33} in appendix \ref{A5}.

We were not able to find the analytic continuation $n_e\to 1$ of the replica; the analytic continuation of replica to non integer values still remains an open problem in the case of disjoint intervals for free compact boson of arbitrary compatification radius.
 
In this work, we numerically checked the free compact boson result against the tight binding model, it would be interesting to check these results against the analytic results for the free massless Dirac field. The symmetry decomposition of the R\'enyi negativity for free massless Dirac field may be obtained using the method in ref. \cite{a30}. It is also important to check the analytical results obtained here with the numerical results obtained for spin chains. Finally, we mention that the holographic dual of the charged R\'enyi moments obtained here could prove to be a useful quantity for determining the symmetry decomposition of negativity in $\text{AdS}_3/\text{CFT}_2$ correspondence. The symmetry resolved entanglement entropy in this context has already been determined in ref. \cite{a36} using the holographic dual of the charged moments.
\appendix
\section{Normalised holomorphic differentials and Period matrix} \label{A1}
In this appendix, we first discuss some formalism in the theory of Riemann surfaces relevant to our present program. We then compute the normalised holomorphic differentials used in the main text and finally, show that the period matrix obtained using our conventions is the same as in eq.\eqref{eq:4.10}.

We first start by defining the conformal transformation which fixes the points $u_1\to 0$, $v_2\to 0$ and $u_2\to\infty$. Let this conformal transformation be a map from $w\to z$ complex plane
\begin{equation} \label{eq:A.1}
z=\frac{(w-u_1)(u_2-v_2)}{(u_2-w)(v_2-u_1)}.
\end{equation}
As discussed earlier we have the n-sheeted Riemann surface $\Sigma_n(x)$ with branch points $x$, $0$, $1$, and $\infty$ and this surface may be parametrised by the elliptic curve in eq.\eqref{eq:4.14}. We now introduce some formalism used in the theory of Riemann surfaces \cite{a46,a47}. The classification of the Riemann surfaces endowed with a conformal structure is achieved by finding an appropriate parametrisation of the associated moduli space, one way is to parametrise the moduli space with the Riemann period matrix $\tau$. For a Riemann surface of genus $g$ the period matrix $\tau$ is a $g\times g$ symmetric matrix.

The period matrix may be given in a cannonical homology basis on the Riemann surface. The basis considered in such problems are two sets of $g$ non contractible loops denoted $a_r$ and $b_r$, and are often called $a$ and $b$ cycles. They satisfy the intersection rules $a_r\circ a_s=b_r\circ b_s=0$ and $a_r\circ b_s=1$. Once this basis is constructed, we may define two $g\times g$ matrices $\mathcal{A}$ and $\mathcal{B}$
\begin{align} 
\mathcal{A}_{rs}=\oint_{a_r}\mathrm{d}z\,w_s(z), \label{eq:A.2}\\
\mathcal{B}_{rs}=\oint_{a_r}\mathrm{d}z\,w_s(z), \label{eq:A.3}
\end{align}
where $w_s(z)$ are the basis of holomorphic differentials (also known as abelian differentials of the first kind). We then proceed to define the basis for normalised holomorphic differentials $\nu_r(z)$ by
\begin{equation} \label{eq:A.4}
\nu_r(z)=w_q(z)\left(\mathcal{A}^{-1}\right)_{qr}.
\end{equation}
These holomorphic differentials satisfy
\begin{align}
\oint_{a_r}\mathrm{d}z\,\nu_s(z)=\delta_{r,s},\label{eq:A.5}\\
\oint_{b_r}\mathrm{d}z\,\nu_s(z)=\tau_{rs}, \label{eq:A.6}
\end{align}
where eq.\eqref{eq:A.5} is the condition of normalisation and $\tau=\mathcal{A}^{-1}\cdot\mathcal{B}=\mathcal{R}+i\mathcal{I}$ is the desired period matrix.
\begin{figure}
\centering 
\includegraphics[width=0.8\textwidth]{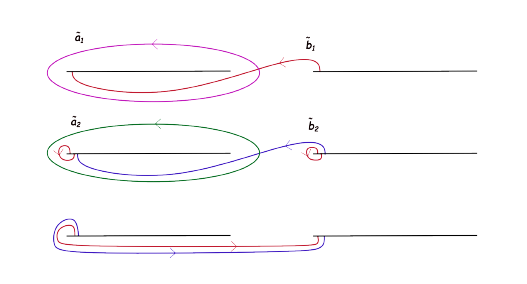}
\caption{\label{fig:viii} Auxiliary homology basis for the Riemann surface $\Sigma_x$ for the case $n=3$. Horizontal black lines are branch cuts on the complex sheet. Left cut is $(x,0)$ and the right cut is $(1,\infty)$. We have depicted 3 complex sheets with each sheet represented by the horizontal couple of branch cut and coloured curves are the Auxiliary homology basis.}
\end{figure}

Returning to our case, the Riemann surface $\Sigma_n(x)$ has genus $(n-1)$. Following ref. \cite{a49} we first choose a set of auxiliary homology basis $\tilde{a}_r$ and $\tilde{b}_r$ (see figure \ref{fig:viii}). Here we mention that this basis is also a cannonical homology basis and similar basis has been used in ref. \cite{a42} to determine the period matrix for Riemann surfaces. However, part of our present aim is to reproduce the period matrix in eq.\eqref{eq:4.10}. So, we choose the homology basis
\begin{equation} \label{eq:A.7}
a_s=\sum_{r=1}^s \tilde{a}_r,\qquad b_s=\tilde{b}_{s}-\tilde{b}_{s+1}\qquad s=1,2,\cdots,n-1,
\end{equation} 
with $\tilde{b}_n=0$. Basis of the holomorphic differentials is given by
\begin{equation} \label{eq:A.8}
w_r(z)=(z(z-1))^{-r/n}(z-x)^{-1+r/n}.
\end{equation}
The auxiliary matrix $\mathcal{\tilde{A}}$ for the loops $\tilde{a}_r$ are computed using eq.\eqref{eq:A.2}
\begin{equation} \label{eq:A.9}
\mathcal{\tilde{A}}_{rs}=i2\pi e^{\frac{i\pi(2r-3)s}{n}}F_{s/n}(x),
\end{equation}
where we have used the relation $\mathcal{\tilde{A}}_{rs}=e^{\frac{i2\pi(r-1)s}{n}}\mathcal{\tilde{A}}_{1s}$. Care must be taken in choosing the appropriate branch for the integral, we have used
\begin{equation} \label{eq:A.10}
\int_0^{x}\mathrm{d}t\,w_r(t+i0^{-})=\frac{\pi}{\sin\left(\pi r/n\right)}F_{r/n}(x).
\end{equation}
Similarly, the auxiliary matrix $\mathcal{\tilde{B}}$ for the loops $\tilde{b}_r$ are computed using eq.\eqref{eq:A.3}
\begin{equation} \label{eq:A.11}
\mathcal{\tilde{B}}_{rs}=-i2\pi e^{\frac{i\pi(r-3)s}{n}}\frac{\sin\left(\pi rs/n\right)}{\sin\left(\pi s/n\right)}F_{s/n}(1-x),
\end{equation}
where we used the integral
\begin{equation} \label{eq:A.12}
\int_x^{0}\mathrm{d}t\,w_r(t+i0^{-})+\int_0^{1}\mathrm{d}t\,w_r(t)=e^{-\frac{i\pi r}{n}}\frac{\pi}{\sin{\left(\frac{\pi r}{n}\right)}}F_{r/n}(1-x).
\end{equation}
Here we mention that we could also arrive at these results by taking the analytic continuation of $x \in(0,1)$ to the complex plane. The hypergeometric function $F_{k/n}(z)$ for complex $z$ is defined on the complex sheet with a branch cut from $1$ to $\infty$. To select the principal branch for $F_{k/n}(1-x)$ for $x\in (-\infty,0)$ we must approach the negative real line from the lower half plane and hence resulting in our choice eq.\eqref{eq:A.10} and eq.\eqref{eq:A.12}.

Using the eq.\eqref{eq:A.9}, and eq.\eqref{eq:A.11} in eq.\eqref{eq:A.7} the matrices $\mathcal{A}$ and $\mathcal{B}$ are readily obtained
\begin{align}
\mathcal{A}_{rs}&=i2\pi e^{\frac{i\pi s(r-2)}{n}}\frac{\sin\left(\frac{\pi rs}{n}\right)}{\sin\left(\frac{\pi s}{n}\right)}F_{s/n}(x),\label{eq:A.13}\\
\mathcal{B}_{rs}&=i2\pi \frac{e^{\frac{i\pi s(r-2)}{n}}}{\sin\left(\frac{\pi s}{n}\right)}\left(\sin\left(\frac{\pi s(r+1)}{n}\right)-e^{-\frac{i\pi s}{n}}\sin\left(\frac{\pi s}{n}\right)\right) F_{s/n}(1-x).\label{eq:A.14}
\end{align}
The normalised holomorphic differentials are then obtained after inverting $\mathcal{A}$ and using eq.\eqref{eq:A.4} 
\begin{equation} \label{eq:A.15}
\nu_r(z)=\sum_{l=1}^{n-1}\frac{e^{-\frac{i2\pi l(r-1)}{n}}\sin{\left(\frac{\pi l}{n}\right)}}{\pi n F_{l/n}(x)}(z(z-1))^{-l/n}(z-x)^{-1+l/n}.
\end{equation}
Finally, the period matrix $\tau$ is computed by using eq.\eqref{eq:A.15} in eq.\eqref{eq:A.6} and this evaluates to eq.\eqref{eq:4.10}.
\section{Cross ratio and Prime forms} \label{A2}
In this appendix, we numerically verify the conjectured relation in eq.\eqref{eq:4.22}, eq.\eqref{eq:4.23}, and eq.\eqref{eq:4.24}. These relations were proved for $n= 2$ and conjectured for general $n$ when $x\in (0,1)$ in ref. \cite{a30}. The proof holds for $x<0$ as well and here we show the proof of the cross ratio function relation for $n=2$ case for the sake of presentation.

For the case $n=2$, the Riemann-Siegel theta functions are just the Jacobi theta functions $\vartheta$. In particular we have the following relations
\begin{align} 
\Theta_{\mathbf{\frac{1}{2}}}(\boldsymbol{w}(0)-\boldsymbol{w}(1)|\tau)&=-ie^{-\frac{i\pi}{4}}e^{-i\frac{\tau}{4}}\vartheta_3(0|\tau), \label{eq:B.1}\\
\Theta_{\mathbf{\frac{1}{2}}}(\boldsymbol{w}(0)-\boldsymbol{w}(\infty)|\tau)&=ie^{-i\frac{\tau}{4}}\vartheta_4(0|\tau),\label{eq:B.2}\\
\Theta_{\mathbf{\frac{1}{2}}}(\boldsymbol{w}(x)-\boldsymbol{w}(1)|\tau)&=-ie^{-i\frac{\tau}{4}}\vartheta_4(0|\tau),\label{eq:B.3}\\
\Theta_{\mathbf{\frac{1}{2}}}(\boldsymbol{w}(0)-\boldsymbol{w}(\infty)|\tau)&=-ie^{-i\frac{\tau}{4}}\vartheta_3(0|\tau),\label{eq:B.4}
\end{align}
where we have used the full period and half period relations for $\theta$ functions given in \cite{a50}. Finally, using the identity $\vartheta_4(0|\tau)/\vartheta_3(0|\tau)=(1-x)^{1/4}$ we have from eq.\eqref{eq:4.24i}
\begin{equation} \label{eq:B.5}
|p(x,0,1,\infty)|=|1-x|^{\frac{1}{2}} .
\end{equation}
\begin{figure}[t]
\centering 
\includegraphics[width=1\textwidth]{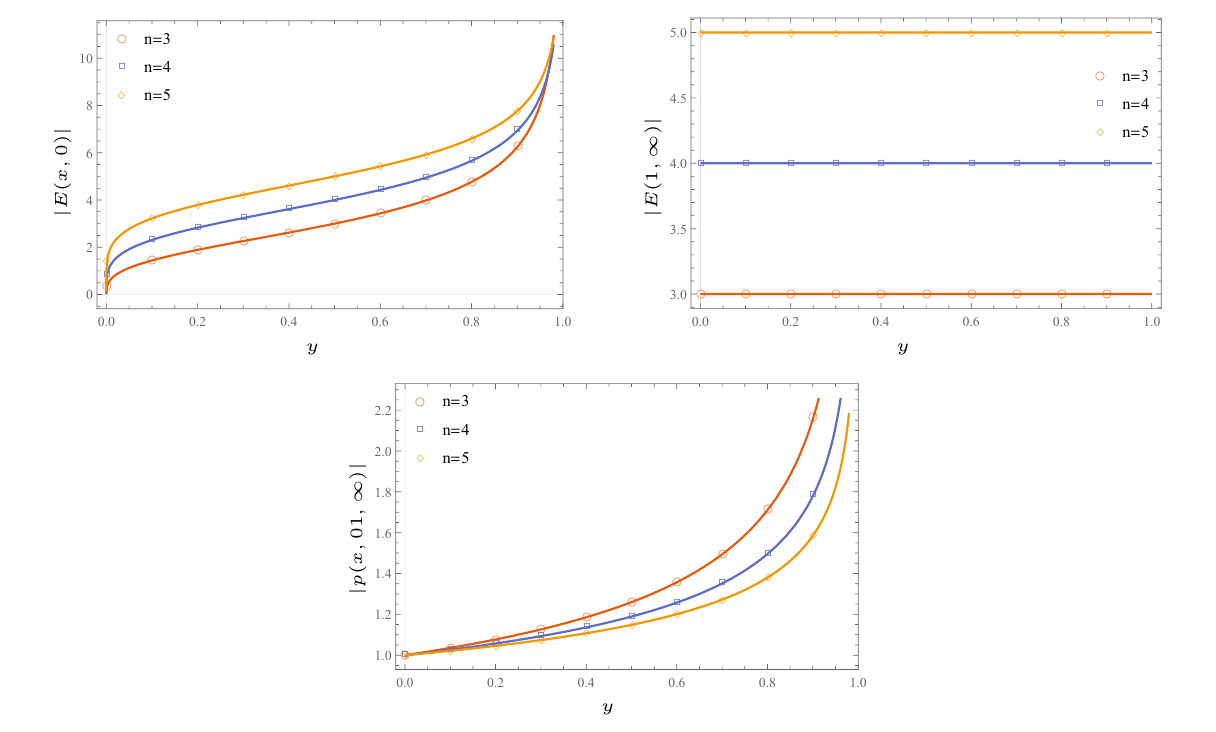}
\caption{\label{fig:ix} Plots for the conjectures in eq.\eqref{eq:4.22}, eq.\eqref{eq:4.23}, and eq.\eqref{eq:4.24}. Horizontal axis is $y=\frac{x}{x-1}$. For vertical axis the solid lines are conjectured functions and the plotted points are numerically computed values of modified prime forms and cross ratio function.}
\end{figure}
In figure \ref{fig:ix}, we show the numerical check for the conjectures in eq.\eqref{eq:4.22}, eq.\eqref{eq:4.23}, and eq.\eqref{eq:4.24} for $n=3,4,5$. For presentation purposes we use the change of variable $x=\frac{y}{y-1}$ and the plot is for $y\in(0,1)$. The numerical results match the analytical results with a good numerical precision.
\section{Numerical model} \label{A4}
In this appendix, we describe the numerical model, namely the tight-binding model or free fermions on a lattice. We used this model to perform numerical checks against the analytical results obtained in section \ref{section4} and section \ref{section5}.

The tight-binding model is described by the Hamiltonian $H=\sum_i \hat{c}^{\dagger}_{i+1}\hat{c}^{\dagger}_{i}+\hat{c}^{\dagger}_{i}\hat{c}^{\dagger}_{i+1}$, where $\hat{c}_i$ is the fermion operator satisfying the anti commutation relation $\{\hat{c}_i,\hat{c}^{\dagger}_j\}=\delta_{i,j}$. The correlation function $C_{ij}=\left\langle\hat{c}^{\dagger}_i\hat{c}_j \right\rangle$, is given by
\begin{equation} \label{eq:C.1}
C_{ij}=\frac{\sin{\left((i-j)\pi/2\right)}}{(i-j)\pi}.
\end{equation}
The partially transposed density matrix for the ground state in the case of two intervals is not a Gaussian matrix. However, it may be expressed as the sum of two Gaussian matrices \cite{a51,a52}
\begin{equation} \label{eq:C.2}
\rho^{T_2}=\frac{e^{-i\pi/4}}{\sqrt{2}}\frac{O_+}{\mathrm{Tr}O_+} +\frac{e^{i\pi/4}}{\sqrt{2}}\frac{O_-}{\mathrm{Tr}O_-},
\end{equation}
where $O_{\pm}=e^{\hat{c}^{\dagger}_{i}W^{\pm}_{ij}\hat{c}_{j}}$ are Gaussian matrices. $W^{\pm}$ are related to the fermion correlation function via the relation $W^{\pm}=\frac{1+G_{\pm}}{1-G_{\pm}}$, where
\begin{equation} \label{eq:C.3}
G_{\pm}=
\left[
\begin{array}{ll}
G_{11} & \pm i G_{12}\\
\pm i G_{21} & -G_{22}
\end{array}
\right],
\end{equation}
with $G_{IJ}=2C_{IJ}-\mathbb{I}$. $C_{IJ}$ are the correlations between the sites in $A_{I}$ with the sites in $A_{J}$. 

The tight binding model has a $U(1)$ symmetry and the corresponding charge operator is $Q=\sum_i\hat{c}^{\dagger}_{i}\hat{c}^{\dagger}_{i}$. In the fermion lattice basis $Q^T_{A_2}=\ell_2-Q_{A_2}$, where $\ell_2$ is the number of lattice sites in $A_2$. The operator $e^{i\alpha\hat{Q}_{A_1}-i\beta\hat{Q}^T_{A_2}}=e^{-i\beta\ell_2}e^{i\alpha\hat{Q}_{A_1}+i\beta\hat{Q}_{A_2}}$ and $e^{i\alpha\hat{Q}_{A_1}+i\beta\hat{Q}_{A_2}}$ may be expressed as a Gaussian matrix with the correlation matrix \cite{a29}
\begin{equation} \label{eq:C.4}
B_{ij}=\left\{
\begin{array}{ll}
\delta_{ij}\frac{e^{i\alpha}}{e^{i\alpha}+1} &\qquad i\in A_1, \vspace{0.1in} \\
\delta_{ij}\frac{e^{i\beta}}{e^{i\beta}+1} & \qquad i\in A_2.
\end{array}
\right.
\end{equation}
Now using the rules of the Gaussian matrix multiplication, the multi-charged moments $R_1^{A_1:A_2}(\alpha,\beta)=\mathrm{Tr}\left(\rho^{T_2}e^{i\alpha\hat{Q}_{A_1}+i\beta\hat{Q}_{A_2}}\right)$ are given by
\begin{equation} \label{eq:C.5}
\begin{split}
R_1^{A_1:A_2}&(\alpha,\beta)=\\
&\frac{(e^{i\alpha/2}+e^{-i\alpha/2})^{\ell_1}(e^{i\beta/2}+e^{-i\beta/2})^{\ell_2}}{\sqrt{2}}\left(e^{-i\pi/4}\det\left(\frac{1-U_+}{2}\right)+e^{i\pi/4}\det\left(\frac{1-U_-}{2}\right)\right),
\end{split}
\end{equation}
where the matrices $U_{\pm}=G_{\pm}(2B-\mathbb{I})$. The charged moment $R_1(\alpha)$ is simply obtained by using $\alpha=\beta$ in eq. \eqref{eq:C.5}. Similarly, for the $n=2$ case, the charged moments $R_2(\alpha)=\mathrm{Tr}\left(\left(\rho^{T_2}\right)^2e^{i\alpha(\hat{Q}_{A_1}+\hat{Q}_{A_2})}\right)$ after some simplification are given by 
\begin{equation} \label{eq:C.6}
R_2(\alpha)=e^{-i\alpha(\ell_1+\ell_2)/2}\det\left(\frac{(1-G_+)(1-G_-)}{4}+e^{i\alpha}\frac{(1+G_+)(1+G_-)}{4}\right).
\end{equation}
The charge imbalance resolved R\'enyi negativity is then given by
\begin{equation} \label{eq:C.7}
\mathcal{R}_n(q)=\frac{1}{2\pi}\int_{-\pi}^{\pi}\mathrm{d}\alpha e^{-i\alpha q-i\alpha\ell_2}R_n(\alpha).
\end{equation}
Notice that we have absorbed the factor $e^{-i\alpha\ell_2}$ in the Fourier transform as it is just a shift in mean charge of the charge imbalance operator $\mathcal{Q}=Q_{A_1}-Q^{T}_{A_2}$.
\section{Revisiting charged moments for complex boson} \label{A5}
In this appendix, we briefly revisit the charged moments for the compact complex boson computed in ref. \cite{a33} and give a correction to the results obtained in the reference.

In the reference, the twist operator $\mathcal{T}_n$ and the flux generating operator $\mathcal{V}_\alpha$ associated with the $U(1)$ symmetry are fused together to give modified twist operators $\mathcal{T}_{n,\alpha}$. This yields a fluxed riemann surface. The partition function on the fluxed riemann surface or the charged R\'enyi moments may be written as $Z_n(\alpha)=Z^{qu}_nZ^{cl}_n$. The quantity $Z^{qu}_n$ is the quantum part given in the reference and $Z^{cl}_n$ is computed to be
\begin{equation}
Z^{cl}_n=\Theta(\mathbf{0}|\eta\Omega)\Theta(\mathbf{0}|\eta\tilde{\Omega}),
\end{equation} 
where $\eta=R^2/2$. $\Omega$ and $\tilde{\Omega}$ are $2n\times 2n$ matrices given in terms of $n\times n$ matrices $U$, $V$, $\tilde{U}$, and $\tilde{V}$ as follows
\begin{equation}
\Omega=\left[
\begin{array}{ll}
U & V\\
V^T & U
\end{array}
\right], \qquad
\tilde{\Omega}=\left[
\begin{array}{ll}
\tilde{U} & \tilde{V}\\
\tilde{V}^T & \tilde{U}
\end{array}
\right].
\end{equation}
The components of the matrices are given by
\begin{align}
U_{rs}&=\frac{i2}{n}\sum_{k=0}^{n-1}\sin\left[\pi a_k\right]\beta_{a_k}(x)\cos\left[\frac{2\pi k(r-s)}{n}\right],\\
V_{rs}&=\frac{i2}{n}\sum_{k=0}^{n-1}\sin\left[\pi a_k\right]\beta_{a_k}(x)\sin\left[\frac{2\pi k(r-s)}{n}\right],\\
\tilde{U}_{rs}&=\frac{i2}{n}\sum_{k=0}^{n-1}\sin\left[\pi a_k\right]\frac{1}{\beta_{a_k}(x)}\cos\left[\frac{2\pi k(r-s)}{n}\right],\\
\tilde{V}_{rs}&=\frac{i2}{n}\sum_{k=0}^{n-1}\sin\left[\pi a_k\right]\frac{1}{\beta_{a_k}(x)}\sin\left[\frac{2\pi k(r-s)}{n}\right],
\end{align}
where $a_k=\left(\frac{k}{n}+\frac{\alpha}{2\pi n}\right)$, $\beta_{a_k}(x)=F_{a_k}(1-x)/F_{a_k}(x)$ and $x\in (0,1)$.
We checked that the matrices $\Omega$ and $\tilde{\Omega}$ are symmetric with positive imaginary part. In the reference $V$, and $\tilde{V}$  were taken to be zero. However this is not true in presence of the flux $\alpha$. Since now the terms in $k$ summation are not odd under $k/n\to 1-k/n$ as opposed to ref. \cite{a41} and components of $V$, and $\tilde{V}$ doesn't sum to zero. Similarly for the charged R\'enyi negativity moments we have the classical contribution to the partition function
\begin{equation}
Z^{cl}_n=\Theta\left(\mathbf{0}|\eta T\right),
\end{equation} 
where $T$ is a $4n\times 4n$ symmetric matrix given in terms of $2n\times 2n$ matrices $A$, $B$, and $W$
\begin{equation}
T=i2\left[
\begin{array}{ll}
A & W \\
W^{T} & B
\end{array}
\right].
\end{equation}
We further write the matrices $A$, $B$, and $W$ in terms of $n\times n$ blocks. These blocks are given by
\begin{align}
\left(A_{11}\right)_{rs}&=\left(A_{22}\right)_{rs}=\frac{1}{n}\sum_{k=0}^{n-1}\frac{\left|\tau_{a_k}\right|^2}{\beta_{a_k}}\sin\left[\pi a_k\right]\cos\left[\frac{2\pi k(r-s)}{n}\right],\\
\left(A_{12}\right)_{rs}&=\left(A_{21}^T\right)_{rs}=\frac{1}{n}\sum_{k=0}^{n-1}\frac{\left|\tau_{a_k}\right|^2}{\beta_{a_k}}\sin\left[\pi a_k\right]\sin\left[\frac{2\pi k(r-s)}{n}\right],\\
\left(B_{11}\right)_{rs}&=\left(B_{22}\right)_{rs}=\frac{1}{n}\sum_{k=0}^{n-1}\frac{1}{\beta_{a_k}}\sin\left[\pi a_k\right]\cos\left[\frac{2\pi k(r-s)}{n}\right],\\
\left(B_{12}\right)_{rs}&=\left(B_{21}^T\right)_{rs}=\frac{1}{n}\sum_{k=0}^{n-1}\frac{1}{\beta_{a_k}}\sin\left[\pi a_k\right]\sin\left[\frac{2\pi k(r-s)}{n}\right],\\
\left(W_{11}\right)_{rs}&=\left(W_{22}\right)_{rs}=\frac{1}{n}\sum_{k=0}^{n-1}\frac{\alpha_{a_k}}{\beta_{a_k}}\sin\left[\pi a_k\right]\sin\left[\frac{2\pi k(r-s)}{n}+\pi a_k\right],\\
\left(W_{12}\right)_{rs}&=-\left(W_{21}\right)_{rs}=\frac{1}{n}\sum_{k=0}^{n-1}\frac{\alpha_{a_k}}{\beta_{a_k}}\sin\left[\pi a_k\right]\cos\left[\frac{2\pi k(r-s)}{n}+ \pi a_k\right],
\end{align}
where $\tau_{a_k}=\alpha_{a_k}+i\beta_{a_k}=iF_{a_k}(1-x)/F_{a_k}(x)$ and $x\in(-\infty,0)$.
\acknowledgments
HG is supported by the Prime Minister’s Research Fellowship offered by the Ministry of
Education, Govt. of India. UY is supported by an Institute Chair Professorship.

\bibliographystyle{JHEP}
\bibliography{main}


\end{document}